\documentclass[9pt,twocolumn,twoside]{optica}
\setboolean{shortarticle}{false}
\setboolean{minireview}{false}

\title{A monolithic 56 Gb/s silicon photonic pulse-amplitude modulation transmitter}

\author{Chi Xiong*, Douglas M. Gill, Jonathan E. Proesel, Jason S. Orcutt, Wilfried Haensch, William M.J. Green}

\affil{IBM T.J. Watson Research Center, 1101 Kitchawan Road, Yorktown Heights, New York 10598, USA}

\affil[*]{Corresponding author: cxiong@us.ibm.com}

\dates{Compiled \today}

\ociscodes{(200.4650) Optical interconnects; (130.4110) Modulators; (060.2330) Fiber optics communications.}

\doi{\url{http://dx.doi.org/10.1364/optica.XX.XXXXXX}}

\begin{abstract}
Silicon photonics promises to address the challenges for next-generation short-reach optical interconnects. Growing bandwidth demand in hyper-scale data centers and high-performance computing motivates the development of faster and more-efficient silicon photonics links. While it is challenging to raise the serial line rate, further scaling of the data rate can be realized by, for example, increasing the number of parallel fibers, increasing the number of wavelengths per fiber, and using multi-level pulse-amplitude modulation (PAM). Among these approaches, PAM has a unique advantage because it does not require extra lasers or a costly overhaul of optical fiber cablings within the existing infrastructure. Here, we demonstrate the first fully monolithically integrated silicon photonic four-level PAM (PAM-4) transmitter operating at 56 Gb/s and demonstrate error-free transmission (bit-error-rate < $10^{-12}$) up to 50 Gb/s without forward-error correction. The superior PAM-4 waveform is enabled by optimization of silicon traveling-wave modulators and monolithic integration of the CMOS driver circuits. Our results show that monolithic silicon photonics technology is a promising platform for future ultrahigh data rate optical interconnects.
\end{abstract}

\setboolean{displaycopyright}{false}

\begin{document}

\maketitle
\thispagestyle{fancy}
\ifthenelse{\boolean{shortarticle}}{\abscontent}{}

\section{Introduction}

Traditional optical interconnects, implemented using parallel multimode fiber coupled to vertical cavity surface emitting laser arrays, face modal dispersion-induced limitations in satisfying longer reach (>100 m) requirements within massively parallel data center and high-performance computing systems \cite{RN37,RN38}. On the other hand, silicon photonics links offer a scalable solution using wavelength division multiplexing, which can be implemented using compact single-mode silicon photonic modulators \cite{RN14,RN36} and on-chip Ge photodetectors \cite{RN16}. As data centers require ever higher data rates, continued multiplexing of wavelengths can be limited by the cost and power consumption of the additional lasers. It is also increasingly challenging to increase the serial line rate of a non-return-to-zero (NRZ) link because the physical limitation of the electro-optic bandwidth of modulators and photodiodes. An alternative solution is to use multi-level signaling formats such as pulse-amplitude modulation (PAM-$m$), where multiple digital bits per symbol are encoded into $m$ different signal amplitude levels. In PAM-4 modulation, two binary bits are encoded into four signal levels, which therefore doubles the data rate at the same symbol rate compared to conventional NRZ links. 

The PAM-4 modulation format is receiving significant attention because of its relative ease of implementation compared to higher-order modulation. PAM-4 modulation has been explored with traditional electrical links \cite{RN13,RN34,RN35} and directly modulated vertical cavity surface emitting lasers \cite{RN32, RN5}. In the case of directly modulated lasers, the PAM-4 signal applied to the laser is generated by summing two NRZ signals followed by amplification with a linear driver. A similar scheme using an electrically generated PAM-4 signal to drive the silicon photonic modulator has been demonstrated \cite{RN7}. Recently it has been demonstrated that a PAM-4 optical waveform can be created by using a silicon photonic modulator with two electrode segments \cite{RN6} and thus equivalently performing the digital-to-analog conversion (DAC) in the optical domain. Using the photonic DAC eliminates the power consumption associated with the electrical DAC circuits. On the receiver side, the PAM-4 signals can be decoded into two separate NRZ streams using an analog-to-digital converter (ADC). Using this approach, a recent demonstration has shown a bit-error rate of $10^{-9}$ at 25.6 Gbaud using a commercial chip with embedded $2^{31}-1$ bit pseudo-random bit sequence (PRBS) generators and error checkers \cite{RN11}. The driver circuits for these demonstrations, however, are implemented using discrete components or hybrid chips, and the parasitics from the packaging can severely degrade the RF signal quality and limit further scaling of data rate.

Here, we demonstrate a two-segment traveling-wave Mach-Zehnder modulator (MZM) silicon photonic PAM-4 transmitter having monolithically integrated current-mode CMOS driver circuits. Monolithic integration of the optical modulator side-by-side with CMOS driver circuits minimizes interconnect and packaging parasitics between the electronics and the photonic device, and allows for minimal performance degradation. In addition, monolithic integration facilitates precision electrical connections and helps, for example, to systematically design for and minimize synchronization issues between the differential drive signals into the two arms of the MZM. Furthermore, monolithic integration facilitates wafer level electro-optic testing, as well as minimizes packaging steps, which can reduce the overall cost. The reported PAM-4 modulator is optimized for high speed performance up to 28 Gbaud (56 Gb/s). We directly measure the bit-error rate of the PAM-4 signal using a standard commercial bit-error tester without post-processing and demonstrate error-free (BER < $10^{-12}$) PAM-4 transmission at 25 Gbaud with $2^{7}-1$ bit PRBS without forward-error correction.

\section{Device design and characterization}
\subsection{Design of the monolithic transmitter}

\begin{figure}[h]
	\centering
	\includegraphics[width=\linewidth]{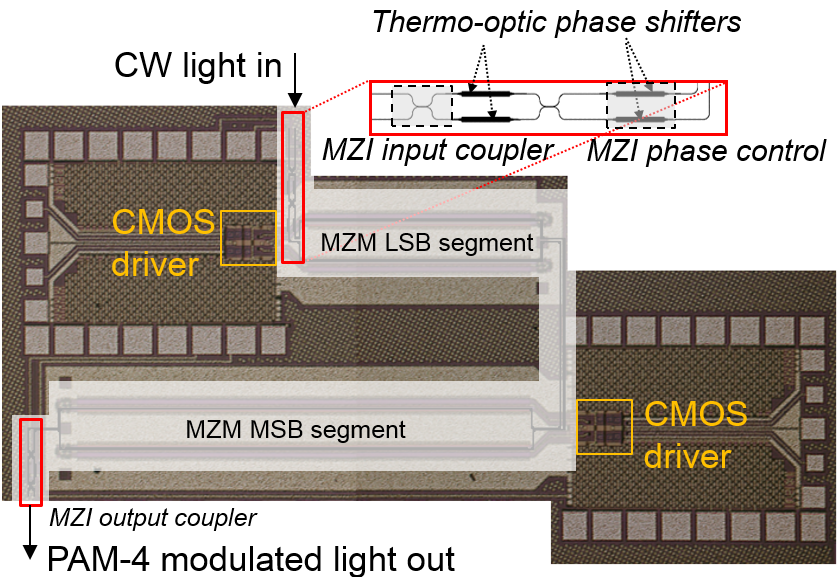}
	\caption{Optical micrograph of the monolithically-integrated PAM-4 transmitter. Two directional couplers with thermo-optically tunable phase shifters serve as input and output couplers for the Mach-Zehnder modulator. The layout of the silicon photonic waveguides is overlaid on top of the micrograph. The magnified waveguide sections in the inset show the thermally-tunable input couplers and MZM phase control. Yellow squares are drawn around the CMOS drivers.}
	\label{fig:micrograph}
\end{figure}

The PAM-4 transmitter is designed and fabricated using IBM’s sub-100nm CMOS integrated nanophotonics technology, CMOS9WG \cite{RN47}. This platform is targeted toward multi-channel short reach O-band (1310 nm) optical interconnects operating at up to 25 Gbaud symbol rates, with the optical components monolithically integrated into the front end CMOS electronics process with a manufacturable low-loss packaging technology \cite{RN29, RN30}. A micrograph of the monolithic PAM-4 transmitter is shown in Fig. \ref{fig:micrograph}. The same CMOS driver has recently enabled an error-free demonstration of a silicon photonic traveling-wave MZM transmitter up to 32 Gb/s \cite{RN46}. The MZM's phase shifter segments have balanced lengths and hence are insensitive to global temperature fluctuations. Each arm of the MZM is divided into two segments, with lengths of 1 mm and 2 mm respectively. Both segments are designed to operate in the traveling-wave mode, and make use of PN junction free carrier depletion-mode electro-optic phase shifters. The Most Significant Bit (MSB) data drives the 2 mm segment and the Least Significant Bit (LSB) drives the 1 mm segment, enabling electro-optic digital-to-analog conversion. Each traveling-wave electrode is terminated by a network of resistors which can be trimmed by focused-ion beam post-fabrication. In this prototype PAM-4 transmitter, this editable termination network provides flexibility to incrementally alter the termination impedance for better matching to the characteristic impedance of the transmission line, which is crucial for optimizing the PAM-4 waveform (see supplementary material). A fixed-value non-trimmable termination resistance can be realized by simply modifying the final metal mask, which is straightforward to implement once all transmission line parameters are finalized in a production-ready design. 

\begin{figure}[h]
	\centering
	\includegraphics[width=\linewidth]{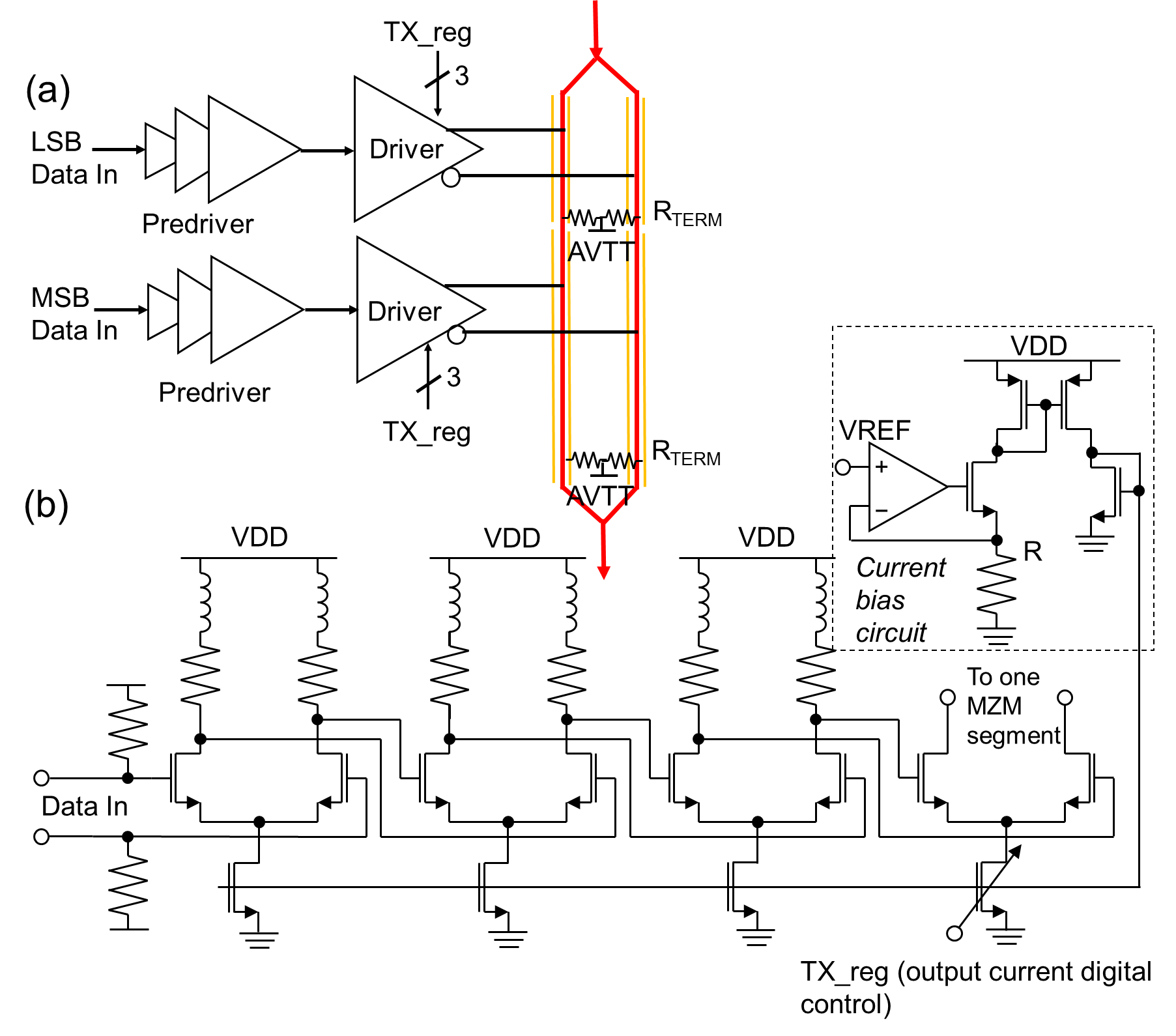}
	\caption{(a) Top-level circuit diagram for the CMOS transmitter driver which consists of a three-stage inductively-peaked current mode logic differential preamplifier followed by a nominal open drain driver. The circuit power supply voltage (VDD) is 1.2 V and the modulator termination resistors R$_{TERM}$ are terminated at a 1.5 V termination power supply (AVTT). The open drain driver’s output amplitude is adjustable using a three-bit digital register (TX\_reg) to offset the MZM’s nonlinear transfer function. (b) Transistor-level circuit schematics of the three-stage predriver and the open drain driver for one MZM segment. The current bias of the driver is set by a reference voltage VREF as shown in the dashed box. With a 30 Ω termination impedance, single-ended output drive swing $V_o$ of the CMOS driver is 1.08 $V_{pp}$ (2.16 $V_{pp}$ differentially).}
	\label{fig:cmos}
\end{figure}

The drivers for MSB and LSB data streams are identical and consist of a three-stage preamplifier followed by a nominal open drain driver. The circuit power supply voltage (VDD) is 1.2 V and the modulator termination resistors R$_{TERM}$ are terminated at a 1.5 V termination power supply (AVTT). The top level circuit diagram is shown in Fig. \ref{fig:cmos}(a). The three preamplifier stages are inductively-peaked current mode logic (CML) differential amplifiers and scale up 2× per stage to drive the open drain driver. With a 30 Ω termination impedance, single-ended output drive swing $V_o$ of the CMOS driver is 1.08 $V_{pp}$ (2.16 $V_{pp}$ differentially). Fig. \ref{fig:cmos}(b) shows transistor level circuit schematics of the three-stage predriver and the open drain driver for one MSM segment. As shown in the dashed box, the current bias circuit creates a bias current proportional to 1/R, so that the voltage drop across all resistors is independent of process variations in sheet resistance. This allows us to maintain a constant voltage swing for CML stages. The current is set by VREF, which establishes the voltage drop across the reference resistor R. The nominal open drain driver’s output amplitude is designed to be adjustable using a three-bit digital control (TX\_reg) to enable a PAM-4 waveform with equal spacing between each symbolic level. In addition, the reference voltage (VREF) can be used to finely adjust the output swing. Operating at 25 Gbaud, VREF of 0.5 V and maximum TX\_reg setting of 7, the VDD supply draws 66 mA current and AVTT draws 36 mA, which amounts to a power consumption of 135 mW. Accordingly, the power consumption is 5.4 pJ/bit at 50 Gb/s and 4.8 pJ/bit at 56 Gb/s.

The traveling-wave MZM is designed for single mode operation at 1310 nm wavelength, and employs PN junction phase shifters embedded within a 145 nm thick silicon-on-insulator (SOI) waveguide layer. The buried oxide beneath the silicon waveguides is 2 µm thick. The phase shifter is based on an interleaved PN junction design similar to that described in \cite{RN10} with 300 nm interdigitated feature size and a nominal peak p and n doping of $3.8\times10^{17}$ cm$^{-3}$. The phase shifter has a measured V$_{\pi}$L of 1.47 V$\cdot$ cm at a PN junction reverse bias of -0.5 V. The CMOS driver differential output is coupled directly into the MZM. Characteristic impedance of the transmission line is extracted from an RF S-parameters test site, which has a design identical to the transmission lines in the MZM in Fig. \ref{fig:micrograph}, but is equipped with input/output RF probe pads. We measure the characteristic impedance of the loaded transmission line to be approximately 30 $\Omega$ across the RF frequency range of interest. 

The MZM has nominal 3 dB directional couplers at both the input and output, each designed using smaller tunable Mach-Zehnder interferometers. Thermo-optic phase shifters, which can be used to offset fabrication drift, are embedded within these directional couplers. In this experiment, the input thermo-optic phase shifter draws 6.5 mA at 1.72 V bias and the output thermo-optic phase shifter draws 1.5 mA at 0.31 V. Accordingly the total power consumption of the input and output thermo-optic phase shifters is 11.6 mW. This power consumption is not intrinsic to the device operation since in the newer generation of devices, we have replaced the thermally tunable directional coupler with wavelength-independent 3 dB couplers \cite{RN98}, which maintain 3 dB splitting ratio over a broad wavelength spectrum without active tuning. An additional thermo-optic phase shifter is used to bias the MZM to the quadrature point of its transfer function. Wafer-level testing results separately show that the propagation loss of the modulator’s PN junction phase shifter waveguide is 10.2 dB/cm.

\begin{figure}[h]
	\centering
	\includegraphics[width=\linewidth]{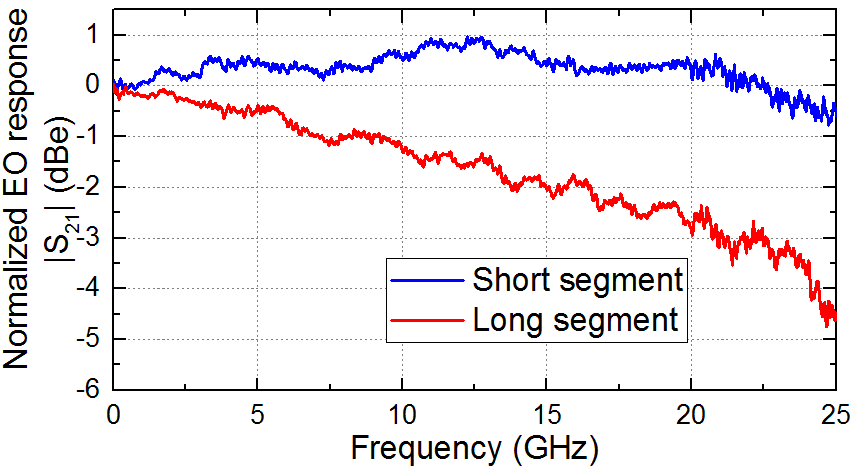}
	\caption{Normalized electro-optic response of the two Mach-Zehnder segments measured on a breakout MZM site without the CMOS drivers. The measurement is taken with a PN junction reverse bias of -0.8 V, which is the DC offset supplied by the CMOS drivers. The long segment has a 3-dB electro-optic bandwidth of 21 GHz. The short segment’s 3-dB bandwidth is beyond the maximum frequency limit of our lightwave component analyzer.}
	\label{fig:bw}
\end{figure}

\subsection{High-speed transmission}

\begin{figure}[h]
	\centering
	\includegraphics[width=\linewidth]{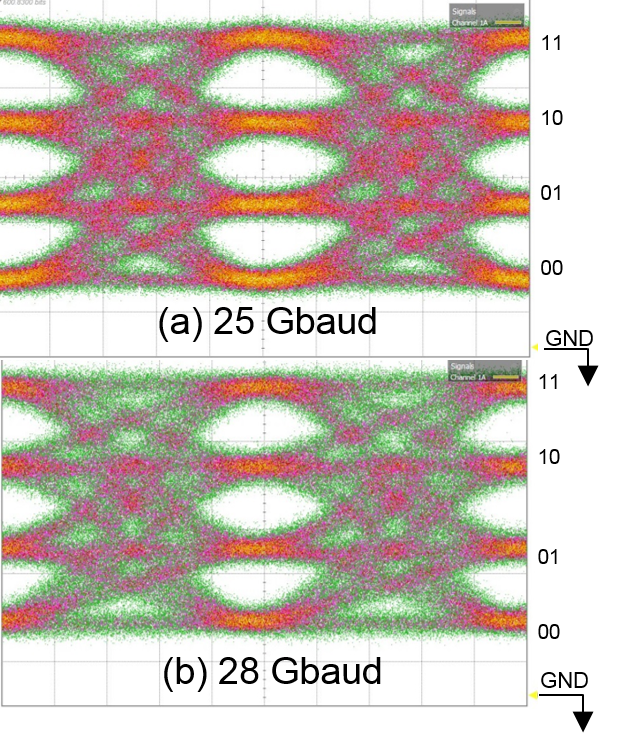}
	\caption{Experimental PAM-4 waveforms at (a) 25 Gbaud (50 Gb/s) and (b) 28 Gbaud (56 Gb/s) measured with $2^{31}-1$ bit PRBS. The extinction ratios between average power level <11> and <00> are measured to be approximately 6.0 dB for both data rates.}
	\label{fig:eyes}
\end{figure}

The electro-optic (EO) response of the two modulator segments is characterized using a breakout PAM-4 modulator site without the CMOS drivers, using a 25 GHz lightwave component analyzer (LCA). The electro-optic measurement has been calibrated up to the tip of the high-speed RF probes. The modulator is adjusted to be at quadrature and a 100 mV$_{pp}$ sinusoidal small-signal was applied to the long and short section of the PAM-4 modulator respectively. The transmitted signal at the output of the photoreceiver inside the LCA is reported in units of electrical power. As shown in Fig. \ref{fig:bw}, the 3-dB electro-optic bandwidth of the long MZM segment is measured to be 21 GHz. The 3-dB bandwidth for the short MZM segment is greater than 25 GHz, which is the bandwidth limit of the analyzer. The long segment has a lower bandwidth because of the larger frequency dependent RF loss associated with the longer loaded transmission line. Following the rule of thumb that the transmitter’s bandwidth needs to be at least 0.75 times the baud rate \cite{RN39}, the 21 GHz bandwidth is sufficient to support 28 Gbaud PAM-4 operation. The termination resistors have been trimmed to be 30 $\Omega$ to match closely with the characteristic impedance of the MZM transmission line, to minimize RF signal reflections as a possible source of signal quality impairment.

Experimentally, to generate PAM-4 optical waveforms, the MSB and LSB data are drawn from the outputs of two non-return-to-zero (NRZ) PRBS pattern generators (Anritsu MP1800A and Centellax SSB16000J) which are triggered by the same external clock. A 1310 nm distributed feedback diode laser with 14 dBm optical output power is coupled onto the chip via a lensed fiber and an on-chip spot-size converter. The typical fiber-to-chip coupling loss per facet is 2.5 dB. An off-chip polarization controller is used to change the polarization of the launched light to transverse electric (TE). Two multi-contact RF probes are used to provide high-speed signals and power to the CMOS circuits. The bit sequences for MSB and LSB data are skewed with respect to each other to reach all possible transitions. High-speed optical eye diagrams at data rates of 25 and 28 Gbaud, shown in Fig. \ref{fig:eyes}, are measured using a 30 GHz bandwidth digital sampling oscilloscope (Agilent DCA-86100D), with both MSB and LSB being driven with $2^{31}-1$ PRBS patterns. Here the extinction ratio (ER) for the PAM-4 waveform is defined as the ratio of the average power level of data <11> to the average power level <00>, ER = 10log$_{10}$(<11>/<00>). The ERs are approximately 6.0 dB at both symbol rates. The eyes show clear openings between all amplitude levels. The measured optical insertion loss of the PAM-4 modulator is 5.0 dB with the CMOS-supplied -0.8 V reverse bias applied to the PN junction phase shifters under PAM-4 operation. 

\subsection{Bit-error rate characterization}

While PAM-4 transmission doubles the data rate, it requires more optical power to achieve the same eye openings (or signal to noise ratio) when compared with the NRZ transmission format. A figure of merit to quantify the transmitter performance is the optical modulation amplitude (OMA) needed to obtain error-free operation, which is usually defined as a bit-error rate (BER) less than $10^{-12}$. For the NRZ waveform the OMA is defined as the difference between the average power level <1> and <0>. For PAM-4 waveform, we focus on the outer OMA (OMA$_{outer}$), which is defined as the difference between the average power level <11> and <00>. The three inner OMA values are given by the difference of high and low power for each eye: OMA$_{01}$=P$_{01}$-P$_{00}$, OMA$_{12}$=P$_{10}$-P$_{01}$, OMA$_{23}$=P$_{11}$-P$_{10}$. The inner OMA, which has the greatest effect on the BER, is defined as the minimum of OMA$_{01}$, OMA$_{12}$, and OMA$_{23}$, i.e., OMA$_{inner}$= min(OMA$_{01}$, OMA$_{12}$, OMA$_{23})$. If the three eye openings have perfect equal distribution, OMA$_{inner}$=1/3$\cdot$ OMA$_{outer}$, i.e. the OMA$_{outer}$ of the PAM-4 waveform will need to be three times the OMA$_{outer}$ of a NRZ waveform at the same symbol rate to yield the same BER. In other words, the theoretical power penalty of a PAM-4 waveform compared to a NRZ waveform at the same symbol rate is 10log$_{10}$3 = 4.8 dB.

To quantify the performance of our PAM-4 transmitter against this theoretical prediction, BER measurements are performed on the PAM-4 transmitter and a reference single-segment Mach-Zehnder NRZ monolithic silicon photonic transmitter with 2.8 mm long phase shifters per arm \cite{RN46}. We first compare the OMA needed to achieve error-free signaling of the PAM-4 and NRZ transmitters at a relatively lower data rate of 12.5 Gbaud. The BER of both transmitters are measured using a 40 Gb/s optical receiver (Discovery Semiconductor R411) and a bit error rate tester (Anritsu MP1800A). The optical power into the receiver is kept low to ensure linear operation and thus an accurate characterization of the multi-level PAM-4 eye diagrams. The measurement of the NRZ BER is straightforward. To measure the PAM-4 BER, however, we treat the three eye openings of the PAM-4 waveforms as three separate NRZ waveforms. and check the BER of the upper (BER$_{upper}$), middle (BER$_{mid}$) and lower eye (BER$_{low}$) against programmed bit patterns: MSB$\land$LSB (Boolean AND between MSB and LSB), MSB, MSB$\lor$LSB (Boolean OR between MSB and LSB) respectively. The aggregate BER of the PAM-4 waveform is then calculated as BER = 1/2$\cdot$BER$_{upper}$ + BER$_{mid}$ + 1/2$\cdot$BER$_{low}$. The BER of the upper and lower eye need to be divided by two because the two bits need to be equally distributed between the upper and lower eye for error checking assuming ‘1’ and ‘0’ bits occur with equal probability in the MSB and LSB streams \cite{RN32}. Because the length of the programmable bit pattern is limited by the memory size of the error detector used, we focus on $2^{7}-1$ PRBS patterns in our experiments. The input MSB and LSB bit streams are phase skewed by half of a word length to ensure full decorrelation. 

\begin{figure}[h]
	\centering
	\includegraphics[width=\linewidth]{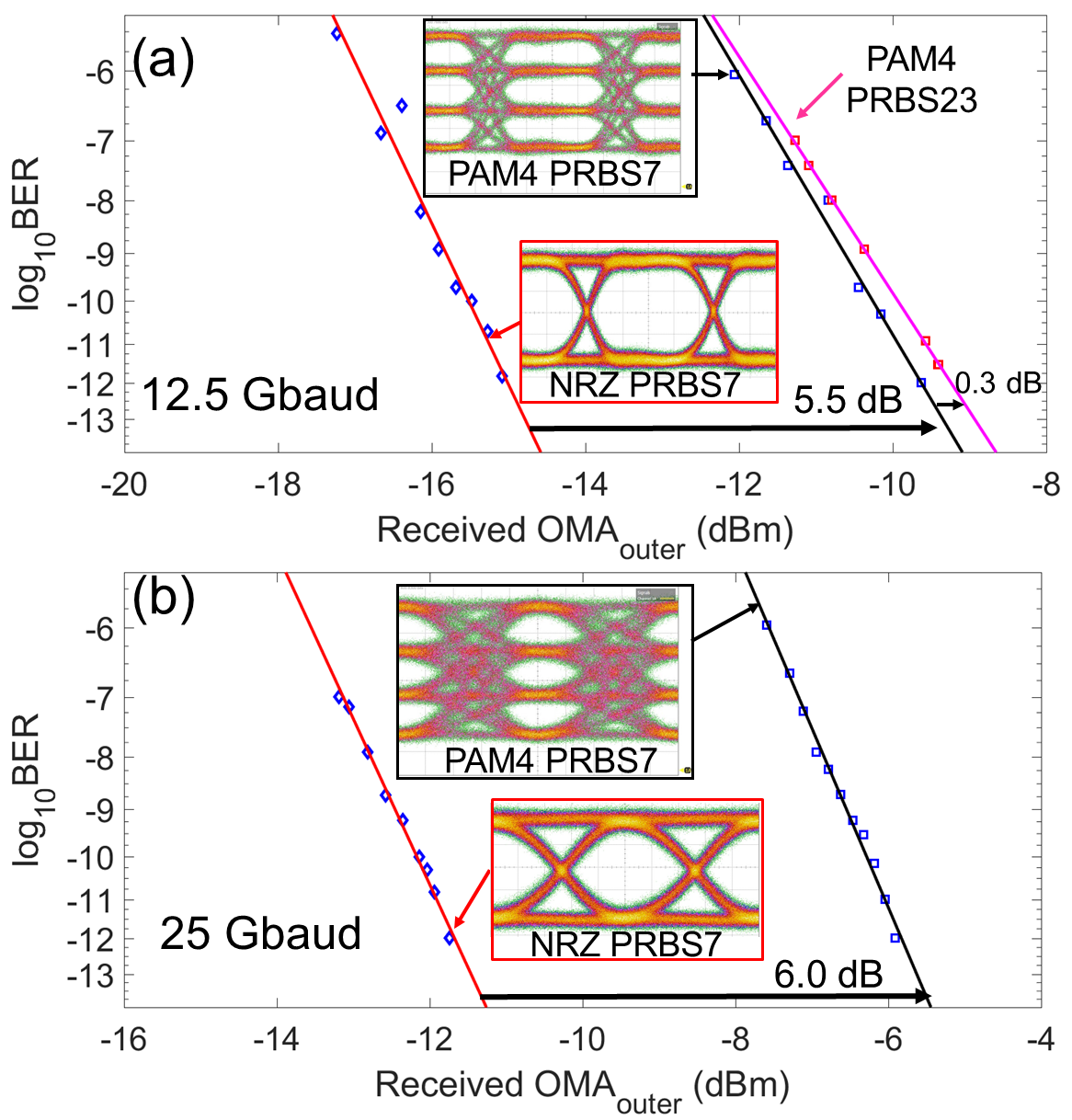}
	\caption{(a) Bit error rate (log$_{10}$BER plotted in logarithmic scale) as a function of OMA$_{outer}$ for PAM-4 and NRZ transmitters operating at 12.5 Gbaud with $2^{7}-1$ bit PRBS patterns. The PAM-4 transmitter shows a 5.5 dB power penalty at BER = $10^{-12}$ relative to the NRZ transmitter. The PAM-4 transmitter tested with $2^{23}-1$ bit PRBS shows an additional 0.3 dB power penalty. The insets show the 12.5 Gbaud PAM-4 and NRZ eye diagrams, which were recorded at a BER < $10^{-12}$. The lines are linear fits to the raw BER data (diamond and circle symbols). (b) BER as a function of OMA for PAM-4 and NRZ transmitters operating at 25 Gbaud with $2^{7}-1$ bit PRBS patterns. The PAM-4 transmitter shows a 6.0 dB power penalty relative to the NRZ transmitter at BER = $10^{-12}$. The insets show the 25 Gbaud PAM-4 and NRZ eye diagrams, which were recorded at a BER < $10^{-12}$. }
	\label{fig:ber}
\end{figure}

As shown in Fig.\ref{fig:ber}(a), the OMA needed to achieve error-free operation (BER<$10^{-12}$) for the 12.5 Gbaud NRZ waveform and PAM4 waveform is -15.3 dBm and -9.8 dBm respectively. In other words, the PAM-4 transmitter shows a 5.5 dB power penalty compared to the NRZ transmitter running at the same symbol rate to achieve equivalent BER. This penalty is thus 0.7 dB greater than the theoretically predicted 4.8 dB power penalty. Part of the excess penalty likely originates from unequal eye openings, because the BER is determined by the smallest eye openings of the three eyes (OMA$_{inner}$). Although the amplitude control of the driver circuit can precisely control the eye spacing, the adjustment is currently done manually which is subject to inaccuracy and drift. In Fig. \ref{fig:ber}(a), the eye with the smallest opening is the upper one, which is measured to have an OMA of 93 percent of  1/3$\cdot$OMA$_{outer}$. As a result, a linearity penalty of ~0.3 dB can be estimated. An output waveform tap and linearity feedback circuit can be implemented in the future to minimize the linearity penalty. The residual 0.4 dB excess penalty can be attributed to PAM-4 implementation penalties as it has been shown that multi-level formats are more susceptible to implementation imperfections including inter-symbol interference due to impedance mismatching, phase skew between MSB and LSB data streams, timing jitter, and so forth \cite{RN32,RN40}.

Longer PRBS patterns have more low frequency spectral content and thus are interesting for characterizing the wide band response of the PAM-4 transmitter. As a result, we measure the BER of the PAM-4 transmitter using a $2^{23}-1$ bit long PRBS pattern, which is the maximum programmable pattern length that can be stored within the 8 Mbit memory of the error detector. A modest additional power penalty of 0.3 dB is measured using $2^{23}-1$ PRBS compared with $2^{7}-1$ PRBS.

The BER for the PAM-4 and NRZ transmitters, both running at 25 Gbaud, is measured using a similar high-speed electro-optic setup. However, achieving error-free operation at 25 Gbaud requires much higher optical power than needed at 12.5 Gbaud. To minimize the nonlinearity at the photoreceiver due to high input optical power, we keep the input optical power at the photoreceiver low while adding a low-noise 40 Gb/s RF amplifier (Picosecond labs 5882) after the photoreceiver to boost the RF signal into the error detector. This effectively improves the input sensitivity of the error detector while minimizing the receiver nonlinearity. As shown in Fig.\ref{fig:ber}(b), the PAM-4 transmitter operates error free (BER<$10^{-12}$) with an OMA greater than -5.8 dBm, with a $2^{7}-1$ bit PRBS pattern. This is the first demonstration of error-free operation of a silicon photonic PAM-4 transmitter at a data rate of 25 Gbaud. We also characterize the reference monolithic silicon photonic NRZ transmitter at 25 Gbaud, and obtain an error-free OMA sensitivity of -11.8 dBm. In other words, the PAM-4 transmitter exhibits a 6.0 dB power penalty relative to the NRZ transmitter at an equivalent BER, with both operating at 25 Gbaud. The excess power penalty compared to theory is 1.2 dB, of which 0.4 dB can be attributed to linearity penalty estimated from the smallest PAM-4 eye opening (middle eye) in Fig.\ref{fig:ber}(b). The remaining 0.8 dB penalty, which is 0.4 dB higher than that at 12.5 Gbaud, could be due to the increased impact of implementation imperfections such as impedance mismatching at 25 Gbaud.

\section{Conclusion}

In conclusion, we have demonstrated a monolithic silicon photonic PAM-4 transmitter in IBM CMOS9WG technology, operating up to 56 Gb/s (28 Gbaud). We directly measure the bit-error rate of the PAM-4 transmitter and compare the relative power penalty with a reference silicon photonic NRZ transmitter. The optimized traveling-wave modulators and monolithic integration of the CMOS drivers enable error-free operation (BER<$10^{-12}$) without forward-error correction at 50 Gb/s with an extinction ratio of 6.0 dB.

For PAM-4 waveform, the optimization of phase alignment between MSB and LSB is important to maximize the horizontal eye opening. In the current experiment, the phase alignment is adjusted using the phase setting in the pattern generators. In future implementation, we can take advantage of the fully monolithic platform to enable a complete on-chip signal distribution network including both the modulator driver and tunable RF delay line on chip \cite{RN99}. To optimize the vertical eye opening ,the monolithic CMOS drivers providing tunable output levels could be used with on-chip germanium power taps and integrated feedback circuitry to actively compensate for drift of the PAM-4 waveform linearity. In addition, the performance of the PAM-4 modulator can be further improved using phase shifter designs with maximized capacitance per unit volume. For example, interleaved junctions with highly scaled pitch have demonstrated smaller V$_{\pi}$L and smaller insertion loss \cite{RN18,RN20}. On a system level, PAM format can be combined with other available techniques for scaling aggregate bandwidth on the silicon photonics platform, including the use of parallel single mode fibers and wavelength division multiplexing, to enable ultrahigh data rate optical interconnects.

\section*{Acknowledgments} The authors acknowledge Marwan Khater, Andreas Stricker, Chris Breslin, Jessie Rosenberg, Edward Kiewra, Tymon Barwicz, Yves Martin, John Ellis-Monaghan, Frederick Anderson, Scott Chilstedt, Michael Nicewicz, Carol Reinholm, Yan Ding, Kate McLean, Michel Paradis, Crystal Hedges, Bruce Porth, Chip Whiting, Mounir Meghelli, Natalie Feilchenfeld, and the rest of the IBM Research and IBM Microelectronics Division (currently GlobalFoundries) teams for their important contributions to this work and the development of CMOS9WG Silicon Photonics technology.


\bigskip \noindent See \href{link}{Supplement 1} for supporting content.



\begin{thebibliography}{10}
	\newcommand{\enquote}[1]{``#1''}
	
	\bibitem{RN37}
	I.~Young, E.~Mohammed, J.~T. Liao, A.~M. Kern, S.~Palermo, B.~Block, M.~R.
	Reshotko, and P.~L. Chang, \enquote{Optical I/O technology for tera-scale
		computing,} Solid-State Circuits, IEEE Journal of \textbf{45}, 235--248
	(2010).
	
	\bibitem{RN38}
	D.~A. Miller, \enquote{Rationale and challenges for optical interconnects to
		electronic chips,} Proceedings of the IEEE \textbf{88}, 728--749 (2000).
	
	\bibitem{RN14}
	G.~T. Reed, G.~Mashanovich, F.~Gardes, and D.~Thomson, \enquote{Silicon optical
		modulators,} Nature Photonics \textbf{4}, 518--526 (2010).
	
	\bibitem{RN36}
	Q.~Xu, B.~Schmidt, S.~Pradhan, and M.~Lipson, \enquote{Micrometre-scale silicon
		electro-optic modulator,} Nature \textbf{435}, 325--327 (2005).
	
	\bibitem{RN16}
	J.~Michel, J.~Liu, and L.~C. Kimerling, \enquote{High-performance Ge-on-Si
		photodetectors,} Nature Photonics \textbf{4}, 527--534 (2010).
	
	\bibitem{RN13}
	A.~Nazemi, H.~Kangmin, B.~Catli, C.~Delong, U.~Singh, T.~He, H.~Zhi, Z.~Bo,
	A.~Momtaz, and C.~Jun, \enquote{A 36Gb/s PAM4 transmitter using an 8b
		18GS/s DAC in 28nm CMOS,} in \enquote{ISSCC. 2005 IEEE International Digest of Technical Papers.
		Solid-State Circuits Conference, 2015.} pp. 1--3.
	
	\bibitem{RN34}
	C.~Menolfi, T.~Toifl, R.~Reutemann, M.~Ruegg, P.~Buchmann, M.~Kossel, T.~Morf,
	and M.~Schmatz, \enquote{A 25Gb/s PAM4 transmitter in 90nm CMOS SOI,} in
	\enquote{ISSCC. 2005 IEEE International Digest of Technical Papers.
		Solid-State Circuits Conference, 2005.},  (2005), pp. 72--73.
	
	\bibitem{RN35}
	A.~Amirkhany, A.~Abbasfar, J.~Savoj, M.~Jeeradit, B.~Garlepp, R.~T. Kollipara,
	V.~Stojanovic, and M.~Horowitz, \enquote{A 24 Gb/s software programmable
		analog multi-tone transmitter,} Solid-State Circuits, IEEE Journal of
	\textbf{43}, 999--1009 (2008).
	
	\bibitem{RN32}
	K.~Szczerba, P.~Westbergh, E.~Agrell, M.~Karlsson, P.~Andrekson, and
	A.~Larsson, \enquote{Comparison of intersymbol interference power penalties
		for OOK and 4-PAM in short-range optical links,} Lightwave Technology,
	Journal of \textbf{31}, 3525--3534 (2013).
	
	\bibitem{RN5}
	R.~Rodes, M.~Mueller, B.~Li, J.~Estaran, J.~B. Jensen, T.~Gruendl,
	M.~Ortsiefer, C.~Neumeyr, J.~Rosskopf, and K.~J. Larsen, \enquote{High-speed
		1550 nm VCSEL data transmission link employing 25 Gbd 4-PAM modulation and
		hard decision forward error correction,} Lightwave Technology, Journal of
	\textbf{31}, 689--695 (2013).
	
	\bibitem{RN7}
	M.~Chagnon, M.~Osman, M.~Poulin, C.~Latrasse, J.-F. Gagné, Y.~Painchaud,
	C.~Paquet, S.~Lessard, and D.~Plant, \enquote{Experimental study of 112 Gb/s
		short reach transmission employing PAM formats and SiP intensity modulator at
		1.3 $\mu$m,} Optics express \textbf{22}, 21018--21036 (2014).
	
	\bibitem{RN6}
	X.~Wu, B.~Dama, P.~Gothoskar, P.~Metz, K.~Shastri, S.~Sunder, J.~Van~der
	Spiegel, Y.~Wang, M.~Webster, and W.~Wilson, \enquote{A 20Gb/s NRZ/PAM-4 1V
		transmitter in 40nm CMOS driving a Si-photonic modulator in 0.13µm CMOS,} in
	\enquote{Solid-State Circuits Conference Digest of Technical Papers (ISSCC),
		2013 IEEE International,}  (IEEE), pp. 128--129.
	
	\bibitem{RN11}
	M.~Traverso, M.~Mazzini, M.~Webster, S.~Anderson, P.-C. Sun, D.~Siadat,
	D.~Conti, A.~Cervasio, S.~Pfnuer, and J.~Stayt, \enquote{25Gbaud PAM-4 error
		free transmission over both single mode fiber and multimode fiber in a QSFP
		form factor based on silicon photonics,} in \enquote{Optical Fiber
		Communication Conference,}  (Optical Society of America), p. Th5B. 3.
	
	\bibitem{RN47}
	N.~B. Feilchenfeld, F.~G. Anderson, T.~Barwicz, S.~Chilstedt, Y.~Ding,
	J.~Ellis-Monaghan, D.~M. Gill, C.~Hedges, J.~Hofrichter, F.~Horst, M.~Khater,
	E.~Kiewra, R.~Leidy, Y.~Martin, K.~McLean, M.~Nicewicz, J.~S. Orcutt,
	B.~Porth, J.~Proesel, C.~Reinholm, J.~C. Rosenberg, W.~D. Sacher, A.~D.
	Stricker, C.~Whiting, C.~Xiong, A.~Agrawal, F.~Baker, C.~W. Baks, B.~Cucci,
	D.~Dang, T.~Doan, F.~Doany, S.~Engelmann, M.~Gordon, E.~Joseph, J.~Maling,
	S.~Shank, X.~Tian, C.~Willets, J.~Ferrario, M.~Meghelli, F.~Libsch,
	B.~Offrein, W.~M.~J. Green, and W.~Haensch, \enquote{An integrated silicon
		photonics technology for o-band datacom,} in \enquote{2015 IEEE International
		Electron Devices Meeting (IEDM),}  (2015), pp. 25.7.1--25.7.4.
	
	\bibitem{RN29}
	T.~Barwicz, Y.~Taira, T.~W. Lichoulas, N.~Boyer, H.~Numata, Y.~Martin, J.-W.
	Nah, S.~Takenobu, A.~Janta-Polczynski, and E.~L. Kimbrell, \enquote{Enabling
		large-scale deployment of photonics through cost-efficient and scalable
		packaging,} in \enquote{Group IV Photonics (GFP), 2015 IEEE 12th
		International Conference on,}  (IEEE), pp. 155--156.
	
	\bibitem{RN30}
	T.~Barwicz and Y.~Taira, \enquote{Low-cost interfacing of fibers to
		nanophotonic waveguides: design for fabrication and assembly tolerances,}
	Photonics Journal, IEEE \textbf{6}, 1--18 (2014).
	
	\bibitem{RN46}
	D.~M. Gill, C.~Xiong, J.~E. Proesel, J.~C. Rosenberg, J.~Orcutt, M.~Khater,
	J.~Ellis-Monaghan, A.~Stricker, E.~Kiewra, Y.~Martin, Y.~Vlasov, W.~Haensch,
	and W.~M.~J. Green, \enquote{Demonstration of error-free 32-Gb/s operation
		from monolithic CMOS nanophotonic transmitters,} IEEE Photonics Technology
	Letters \textbf{28}, 1410--1413 (2016).
	
	\bibitem{RN10}
	J.~Rosenberg, W.~Green, S.~Assefa, D.~Gill, T.~Barwicz, M.~Yang, S.~Shank, and
	Y.~Vlasov, \enquote{A 25 Gbps silicon microring modulator based on an
		interleaved junction,} Optics Express \textbf{20}, 26411--26423 (2012).
	
			\bibitem{RN98}
			K.~Jinguji and N.~Takato and A.~Sugita and M.~Kawachi,
			\enquote{Mach-Zehnder interferometer type optical waveguide coupler with wavelength-flattened coupling ratio,} Electronics Letters \textbf{26}, 1326--1327 (1990).
	
	\bibitem{RN39}
	A.~M.~T. Pedretti, \enquote{Monolithic separate absorption and modulation
		Mach-Zehnder wavelength converters,} Ph.D. Thesis, University of California, Santa Barbara (2007).
	
	\bibitem{RN40}
	S.~Kumar, \emph{Impact of Nonlinearities on Fiber Optic Communications}, vol.~7 (Springer Science \& Business Media, 2011).
		
		\bibitem{RN99}
		P.~Rito, I.~G.~López, D.~Petousi, L. ~Zimmermann, M.~Kroh, S.~Lischke, D.~Knoll, D.~Kissinger and A.~C.~Ulusoy, \enquote{A monolithically integrated segmented driver and modulator in 0.25 $\mu$m SiGe: C BiCMOS with 13 dB extinction ratio at 28 Gb/s,} in \enquote{2016 IEEE MTT-S International Microwave Symposium (IMS),}  (IEEE), pp. 1--4.
		
	\bibitem{RN18}
	I.~Goykhman, B.~Desiatov, S.~Ben-Ezra, J.~Shappir, and U.~Levy,
	\enquote{Optimization of efficiency-loss figure of merit in carrier-depletion
		silicon Mach-Zehnder optical modulator,} Optics Express \textbf{21},
	19518--19529 (2013).
	
	\bibitem{RN20}
	E.~Timurdogan, C.~M. Sorace-Agaskar, E.~S. Hosseini, and M.~R. Watts,
	\enquote{An interior-ridge silicon microring modulator,} Lightwave
	Technology, Journal of \textbf{31}, 3907--3914 (2013).
		
\end{thebibliography}



\end{document}